# Statistical Mechanics of Topological Fluctuations in Glass-Forming Liquids

Katelyn A. Kirchner[1], Seong H. Kim[1,2], and John C. Mauro[1,*]

[1]*Department of Materials Science and Engineering, The Pennsylvania State University, University Park, Pennsylvania 16802, USA*
[2]*Department of Chemical Engineering, The Pennsylvania State University, University Park, Pennsylvania 16802, USA*
*Corresponding Author:* jcm426@psu.edu

**Abstract:** All liquids are topologically disordered materials; however, the degree of disorder can vary as a result of internal fluctuations in structure and topology. These fluctuations depend on both the composition and temperature of the system. Most prior work has considered the mean values of liquid or glass properties, such as the average number of topological degrees of freedom per atom; however, the localized fluctuations in properties also play a key role in governing the macroscopic characteristics of any glass-forming system. This paper proposes a generalized approach for modeling topological fluctuations in glass-forming liquids by linking the statistical mechanics of the disordered structure to topological constraint theory. In doing so we introduce the contributions of localized fluctuations into the calculation of the topological degrees of freedoms in the network. With this approach the full distribution of properties in the disordered network can be calculated as an arbitrary function of composition, temperature, and thermal history (for the nonequilibrium glassy state). The scope of this current investigation focuses on describing topological fluctuations in liquids, concentrating on composition and temperature effects.

**I. Introduction**

Unlike a crystal, where the atomic positions are clearly defined, glass-forming systems have a non-crystalline structure where the details of atomic coordinates and bond configurations must be described in terms of statistical distributions. Due to the role of configurational entropy, liquids have localized fluctuations (or local variations in disorder) inherent in their structure, bonding configurations, and network topology [1]-[4]. These variations are manifested in the broadening of absorption peaks or shapes in spectroscopy [5]-[7]. In multicomponent systems, the fluctuations depend on composition, temperature, and (for the glassy state) thermal history. For example, a hyperquenched glass fiber has a higher entropic contribution to its structure, and hence a greater level of atomic fluctuations, compared to a slowly cooled or annealed sample having the same chemical composition [8]-[9].

Localized fluctuations in glass-forming systems are important for many practical applications. For example:

- The attenuation in low-loss optical fibers is dominated by Rayleigh scattering. Rayleigh scattering is a function of density fluctuations, which scale nonmonotonically with the thermal history of the glass [10]-[11].

- The relaxation behavior in glass is comprised of a variety of different relaxation modes operating over a range of time scales. This spectrum of relaxation modes relates directly to atomic scale fluctuations in the glass structure, with some structural motifs having shorter relaxation times. This behavior is critical for glasses used in high performance display applications [12]-[15].

- Nucleation and phase separation are likely to be governed by localized fluctuations in bonding, since over-constrained regions of the network create localized stresses that can be relieved through crystallization. Compositional inconsistencies can also serve as a precursor to larger scale phase separation [16]-[19].

- Localized fluctuations in bonding and topology play a governing role in the mechanical properties and fracture behavior of glasses. Cracks propagate preferentially through regions where the glass network is under-constrained, since there are fewer bonds to break. As a result, the toughness and damage resistance of glass can potentially be enhanced through tailoring the localized bond fluctuations [20].

Despite the scientific and technological importance of compositional and topological fluctuations in glasses, very few studies have been conducted to elucidate this subject. To date, most studies of glass composition-structure-property relationships have focused primarily on *mean-field* descriptions, averaging over the fluctuation effects and making the connection between average values of structural features and the resulting macroscopic properties of the system [21]-[26]. However, as non-crystalline materials, glass-forming systems have inherent atomic scale fluctuations, i.e., distributions in their structure and bonding. These inconsistencies play a key role in determining many of their thermophysical properties, including the details of their mechanical properties and relaxation behavior [7], [13], [27].

This paper presents a general modeling approach describing structural and topological fluctuations in the metastable equilibrium supercooled liquid state. By linking statistical mechanics with topological constraint theory, we analyze the factors governing modifier speciation. We then present several model calculations demonstrating the effects on the properties of the liquid network, including details of its topological degrees of freedom. The physics of the statistical and topological constraint approaches are linked through the strength of bonding in the glass-forming network. By calculating the probability density associated with the number of atomistic constraints in the network, we describe the structural and bonding characteristics of the localized and network glass-forming system as a function of composition and temperature. Whereas most prior work in the field has focused on the average number of constraints per atom, this paper extends constraint theory by incorporating localized fluctuations in rigidity, consistent with the statistical mechanical model of the underlying atomic structure.

## II. Statistical Mechanics of Glass-Forming Structures

Composition and temperature govern the atomic structure and resulting macroscopic properties of a glass-forming system [28]-[30]. Glass-forming systems are typically composed of network formers, which are atoms that form the main backbone of the covalently connected network, and network modifiers, which are ionically bonded atoms that alter the structure and topology of the network [31]. The presence of modifiers affects the microscopic coordination number of the nearest neighbor network formers, which has a large influence on the macroscopic structural properties of the system [1]. An additional challenge in glass physics is that the system is continuously relaxing toward its metastable equilibrium liquid state [32]-[40]. Statistical mechanical approaches allow us to capture this time dependent relaxation and connect the microscopic and macroscopic physics [41]-[42].

Diffraction experiments and atomistic simulations (e.g., molecular dynamics) have provided information regarding the statistics of glass structure, such as the probability distributions of atomic positions and the time evolution of the material [43]-[45]. While molecular dynamics simulations aid in the understanding of structural and energetic properties, such as the energies associated with various types of bond constraints, these simulations are unable to capture all of the material's features due to time and length scale restrictions. Thus, they are too computationally intensive to be applied for thorough compositional studies [46]-[49].

*Ab initio* simulations offer more accurate descriptions but at a higher computational cost, making them infeasible for large systems or long time scales [50]-[54]. In order to circumvent the problems associated with traditional atomic scale simulations, Mauro [1], [3] developed a statistical mechanical model based on a hypergeometric distribution of site occupancies. In this approach, a network modifier (such as an alkali or alkaline earth ion in an oxide glass) occupies a network forming site based on a relative probability. Once an association is made, that site is no longer available for further occupation (i.e., bonding) [1]. This statistical mechanical approach is an effective method for capturing the distribution of structural motifs in a glass-forming network [55].

Statistical mechanics can be used to describe the inherent disorder in glass-forming systems and calculate the macroscopic properties associated with a suitable ensemble of microscopic states [40], [46], [56]-[65]. The approach for implementing statistical theory must be addressed carefully due to the nonequilibrium and nonergodic nature of glass [2], [66]-[69]. Here, we focus on describing the metastable supercooled liquid state, where the fluctuations can be considered in equilibrium, using a partition function that excludes the crystalline state.

According to the statistical mechanical model derived previously [1], [3], the probability of occupying each site is based on competition between entropic and enthalpic contributions. The relative concentration of each network former dictates the entropic preferences. To account for the enthalpic contribution, a Boltzmann weighting factor, $\omega_i$, is incorporated, given by [1], [3]

$$\omega_i = \exp\left(-\frac{\Delta H_i}{kT}\right), \tag{1}$$

where $\Delta H_i$ is the enthalpy change associated with occupying site type $i$, $k$ is Boltzmann's constant, and $T$ is the absolute temperature. In glassy systems the fictive temperature, $T_f$, can substitute for the physical temperature, $T$, as an approximation of the nonequilibrium structure. For the metastable supercooled liquid the free energy is minimized at the given temperature, $T$, for the ensemble of disordered microstates. The weighting factor in Eq. (1) captures this phenomenon to ensure that the minimum free energy is attained; hence, the system is in equilibrium [1].

With this Boltzmann weighting factor, the probability of occupying site type $i$ with the $m^{th}$ modifier atom is given by [1], [3]

$$p_{i,m} = \frac{1}{Q_{m-1}}\left(g_i - a_{i,m-1}\right)\exp\left(-\frac{\Delta H_i}{kT}\right), \tag{2}$$

where $Q_{m-1}$ is the path-dependent partition function calculated after occupation of modifier $m - 1$ (which normalizes the distribution at each step), $g_i$ is the population size of network former sites of type $i$, and $a_{i,m-1}$ is the number of type $i$ sites previously occupied after modifier $m - 1$. Due to the incorporation of the Boltzmann weighting factor, Eq. (2) yields a noncentral hypergeometric distribution of site occupation probabilities. Eq. (2) thus provides a means of calculating the statistics of glass structure as a function of composition and temperature to serve as an input for topological constraint theory [3].

## III. Topological Constraint Theory of Glass-Forming Systems

Topological constraint theory, also known as rigidity theory, analyzes the connectivity of a glass-forming network to determine related properties by comparing the number of atomic degrees of freedoms with the number of interatomic force field constraints [3]. This approach is based on the work of Phillips and Thorpe [56], [70]-[74], and that of Gupta and Cooper [64], [75]-[77], who first proposed treating glass as a network of constraints. By comparing the number of constraints to the number of atomic degrees of freedom, it is possible to predict the behavior of a number of mechanical and thermal properties.

Gupta and Mauro [62], [78] extended this theory by implementing a probability function for constraint rigidity that accounts for thermal effects and varying strengths of different types of bonds. With the Gupta-Mauro *temperature-dependent constraint* theory [62], the various constraints are no longer considered equal, allowing calculation of the number of atomic constraints as a function of both composition and temperature, given by $n(T, x)$ [3], [79]-[83]. $n(T, x)$ is considered to be separable, following the double summation [78]

$$n(T, x) = \sum_i N_i(x) \sum_\alpha w_{i,\alpha} q_\alpha(T) , \qquad (3)$$

where the first sum is over the network-forming species $i$ and the second sum is over the various constraints $\alpha$. $N_i(x)$ is the mole fraction of network-forming species $i$ in composition $x$, $w_{i,\alpha}$ is the number of constraints type $\alpha$ associated with the species type $i$, and $q_\alpha(T)$ is the temperature-dependent rigidity of constraint $\alpha$. By denoting $i$ as a network-forming species, we mean that $i$ is any atom bonded to at least two other atoms within the liquid or glass network [78]. Network topology is responsible for critical behaviors in a number of mechanical and thermal properties; hence, it enables direct calculation of the macroscopic properties of the glass-forming system [46], [84]-[96].

Prior work has treated $N_i(x)$ as the average mole fraction of each species, making $n(T, x)$ the average number of constraints per atom [62], [78]. This paper extends topological constraint theory by incorporating the full probability distributions of $N_i(x)$, as calculated from the statistical mechanical approach in Section II, to quantify the localized fluctuations in rigidity in addition to the mean rigidity of the network.

The microscopic rigidity, $q_\alpha(T)$, is defined as a bond's resistance to bending, stretching, twisting, breaking, or other deformations. Following the temperature-dependent constraint approach, $q_\alpha(T)$ is a unitless scaling factor that ranges from 0 (fully flexible) to 1 (fully rigid). The rigidity for each constraint, $\alpha$, is an independent function of temperature, $T$, given by [66], [78]

$$q_\alpha(T) = \left[\, 1 - \exp\left(-\frac{\Delta F_\alpha^*}{kT}\right) \right]^{vt_{obs}} ,$$ (4)

where $v$ is the vibrational attempt frequency, $t_{obs}$ is the observation time, and $\Delta F_\alpha^*$ is the activation free energy for breaking the $\alpha$ constraint, given by [78]

$$\Delta F_\alpha^* = -kT_\alpha \ln\left(1 - 2^{-\frac{1}{vt_{obs}}}\right) ,$$ (5)

where $T_\alpha$ is the onset temperature for constraint $\alpha$. The onset temperature, $T_\alpha$, is defined as the temperature at which the probability of breaking the constraint is exactly 50%.

Eq. (4) is derived by considering the probability of a particle remaining trapped in a free energy well, as represented in Figure 1. In order to break the constraint, the particle needs only to escape once: if the escape is successful, then the bond is broken. In Eq. (4), $1 - \exp(-\Delta F_\alpha^*/kT)$ is the probability of not escaping in a single attempt, and the exponent, $vt_{obs}$, accounts for the total number of escape attempts, i.e., the total number of vibrations within the observation time period. If the probability of a bond breaking is zero, i.e., $q_\alpha(T) = 1$, then the constraint is fully rigid. When $q_\alpha(T) = 0$, the bond is fully flexible. Values of $q_\alpha(T)$ between zero and one indicate a degree of partial rigidity, meaning a fraction of the constraints (equal to $q_\alpha(T)$) are rigid, while the remaining fraction $(1 - q_\alpha(T))$ are flexible. Figure 2 shows a plot of Eq. (4) to visualize the temperature dependency of rigidity about the onset temperature $T_\alpha$. As $T \to 0$, there is zero thermal energy available to break the constraints, so all constraints are fully rigid ($q_\alpha(T) = 1$). As $T \to \infty$, thermal energy becomes large enough to enable all of the constraints to break easily ($q_\alpha(T) = 0$); hence, the bond is considered flexible. Molecular dynamics simulations have also verified these theoretical results by calculating bond rigidity as a function of temperature [49].

The constraint onset temperature given in Eq. (5) depends on the free energy associated with the constraint, i.e., the relative entropy and enthalpy associated with each type of bond. For consistency between the structural and topological models, these parameters must be considered in concert with the corresponding parameters of the statistical mechanical model described in Section II.

## IV. Linking the Statistical Mechanical Model to Topological Constraint Theory

Both structure and topology are based on competition between entropic and enthalpic contributions to the thermodynamics of the system. Eq. (4) considers the Helmholtz free energy, $\Delta F_\alpha^*$, of the constraint $\alpha$, which is appropriate for an isochoric system; however, most real glass formation is under isobaric conditions. Hence, the Gibbs free energy, $\Delta G_\alpha = \Delta H_\alpha - T\Delta S_\alpha$, can be substituted:

$$q_\alpha(T) = \left[1 - \exp\left[-\frac{\Delta H_\alpha}{kT}\right]\exp\left[\frac{\Delta S_\alpha}{k}\right]\right]^{\nu t_{obs}}. \tag{6}$$

The constraint onset temperature therefore depends on both the enthalpic ($\Delta H_\alpha$) and entropic ($\Delta S_\alpha$) contributions to the transition barrier.

The entropic contribution in the constraint model is a different type of entropy compared to the entropic contribution in the statistical mechanic model. In the statistical model, the entropy is based on the number of sites available for bonding, which can vary dramatically with the composition of the liquid. In the constraint model, the entropy defines the number of pathways along which a bond can break. In real systems the number of vibrational pathways remains approximately constant, making the entropic contribution in Eq. (6) minimal. Hence, in this paper it is assumed that the change in bond entropy is negligible, making the bond enthalpy equal to the Gibbs free energy barrier and allowing us to rewrite Eq. (6) as

$$q_\alpha(T) = \left[1 - \exp\left[-\frac{\Delta H_\alpha}{kT}\right]\right]^{\nu t_{obs}} \quad . \tag{7}$$

The set of relevant bond enthalpies depends on which bonding configurations are possible. In a binary system of A and B network forming sites, there are two available configurations for each site: *unoccupied* (i.e., not bonded to a network modifier, which we represent as $A, un$ and $B, un$) or *occupied* (i.e., the network former is associated with a network modifier, represented as $A, oc$ and $B, oc$).

Recall from Eq. (3) that the number of rigid constraints per atom, $n(T, x)$, is dependent on both the constraint's rigidity and the number (i.e., the count) of each constraint. Therefore, if the modifier occupies site B, site A is left unoccupied. In this case, the constraints that exist are

$$H_{B,oc}w_{B,oc} + H_{A,un}w_{A,un} \quad . \tag{8}$$

Conversely, if site B is occupied, then that modifier association does not have a B unoccupied site or an A occupied site. As such, the system's change in enthalpy between the two possible sets of configurations is given by

$$\Delta H = H_{B,oc}w_{B,oc} + H_{A,un}w_{A,un} - H_{Aoc}w_{A,oc} - H_{B,un}w_{B,un} \quad . \tag{9}$$

Eq. (9) mathematically defines the relationship for the enthalpic contributions of the sites present in the network. Combining Eq. (7) and Eq. (9) gives us the total change in rigidity due to occupying a B site:

$$q_\alpha(T) = \left[1 - \exp\left[-\frac{H_{B,oc}w_{B,oc} + H_{A,un}w_{A,un} - H_{Aoc}w_{A,oc} - H_{B,un}w_{B,un}}{kT}\right]\right]^{\nu t_{obs}}. \qquad (10)$$

To elucidate the interdependence of these various contributions introduced in Sections II - IV, we begin by studying the most simplistic case. Then, each subsequent case adds another layer of complexity to better understand the effects of composition and temperature on the distribution and mean of the number of rigid atomistic constraints. The results for these calculations are presented in Section V.

## V. Results and Discussion

Our model represents a glass-forming liquid with two network formers, A and B, and one type of modifier, M. Each modifier ion has two roles by which it can affect the structure and topology of the glass-forming system [78], [97]-[98]. Figure 3 illustrates the modifier speciation for the case of an alkali borosilicate. The modifier can produce a non-bridging oxygen, which decreases the number of rigid constraints in the network [99]-[103] or the modifier can convert the boron from threefold to fourfold coordination, which increases the number of constraints, thereby increasing the rigidity (the so-called "boron anomaly") [38], [78].

Two initially unoccupied network formers, A and B, have equal mole fractions: $[A] / ([A] + [B]) = [B] / ([A] + [B]) = 0.5$. In this case site A and B are equally favored from an entropic viewpoint, allowing our initial investigation to focus on understanding the effects of their relative enthalpies. Modifiers are then introduced to occupy 50% of the network forming sites: $[M] / ([A] + [B]) = 0.5$. This concentration was chosen so that all modifiers have the option to fully occupy one type of site, if that is thermodynamically preferred. As a result of the network modifier-former associations, the final structure has four types of states. There are two available states for each site; either it is occupied by a modifier, (represented as $A, oc$ and $B, oc$) or unoccupied ($A, un$ and $B, un$). Being unoccupied means the site lacks a modifier association. To accommodate the enthalpic differences, we assume that $H_{A,oc} > H_{B,oc}$, meaning that occupying site B is enthalpically favored over occupying site A.

A stronger bond has a more negative bond energy; hence, if our B site has a lower enthalpy, it is stronger and thus produces more rigid bonds. As a result, occupying site B will increase the rigidity of the constraint. In our investigation of $n(T, x)$ we will first examine the dependence on composition and then investigate the temperature dependence.

## A. Composition Dependence

In determining the dependence on composition ($x$), our model assumes that the $A, un$ and $B, un$ states have a rigidity of 0.5, and $A, oc$ decreases site A's rigidity to 0.25, while $B, oc$ increases site B's rigidity to 0.75; in other words, for the purposes of these calculations we assume $q_{A,un} = q_{B,un} = 0.5$; $q_{A,oc} = 0.25$; $q_{B,oc} = 0.75$.

As the difference in enthalpy between $H_{A,oc}$ and $H_{B,oc}$ increases, site B becomes even more enthalpically preferred compared to site A. Utilizing Eq. (1), Figure 4 maps the effect of $\Delta H$ on the relative mole fraction of sites occupied. When $\Delta H / kT \rightarrow 0$, every site is equally favored; hence, the structure and topology of the system are fully governed by entropy. When $\Delta H / kT \rightarrow \infty$, enthalpy completely governs the modifier speciation, causing site B to be preferred. Real glass systems typically fall between these two extremes, where there is a competition between entropy and enthalpy to dictate which fraction of sites becomes occupied [1]. This competition between entropic and enthalpic factors explains the so-called "mixed network former effect" in glass systems [1].

Recall that the mole fraction of sites, $N_i(x)$, is a factor impacting the number of rigid constraints per atom, as indicated in Eq. (3). Hence, to model the distribution of $n(T, x)$ we need to calculate the distribution of $N_i(x)$, which is captured in Figure 5. Similar to Figure 4, as $\Delta H/kT$ increases the modifiers prefer to occupy site B. As $\Delta H/kT \to 0$, entropy dominates, so the probability density function follows a central hypergeometric distribution. As $\Delta H/kT \to \infty$, enthalpy dominates, and the bonding configuration becomes more ordered, causing the standard deviation to decrease. As the breath of the distribution decreases, each sample at that composition is increasingly homogenous.

Aside from mole fraction of each species, the rigidity of each constraint, $q_\alpha(T)$, and the number of each constraint, $w_{i,\alpha}$, also govern the distribution of $n(T, x)$. Figure 6(a) captures this dependence by mapping the effect of changing the magnitude of a localized constraint. Here $q_{B,oc}$ varies over five different magnitudes of rigidity, while $q_{A,un} = q_{A,oc} = q_{B,un} = 0.5$ are held fixed. When $q_{B,oc} = 0.5$, all sites are equivalent; therefore, each sample is identical, as represented by the Dirac delta function. As the difference in rigidity between each localized constraint increases, the breadth of the distribution of the number of rigid atomistic constraints also increases.

Similar to increasing $q_\alpha(T)$, when $w_{i,\alpha}$ increases, the mean of $n(T, x)$ also increases, as shown in Figure 6(b). Since the number of constraints must be a positive integer, $w_{i,\alpha} = 1$ results in the minimum mean value of $n(T, x)$. In this scenario, each site has the same rigidity ($q_{A,un} = q_{A,oc} = q_{B,un} = q_{B,oc} = 0.5$) and is enthalpically and entropically identical. $w_{A,un} = w_{A,oc} = w_{B,un} = 1$ while $w_{B,oc}$ varies over three levels. Recall from Eq. 3 that $n(T, x)$ is the summation, $\sum_\alpha w_{i,\alpha} q_\alpha(T)$. Therefore, as either $w_{i,\alpha}$ or $q_\alpha(T)$ increases, the mean value of $n(T, x)$ also increases.

The composition of the glass-forming system is perhaps the most important factor influencing its properties. To determine the impact of composition on fluctuations in structure and topology, we consider varying the site B concentration, $[B]$, and modifier concentration, $[M]$, which are shown in Figure 7 and Figure 8, respectively. Figure 7 assumes that $\Delta H/kT = 0$, making this is a fully entropically driven system. Occupying a B site increases the localized, and hence overall, number of rigid constraints within the system. Thus, as B sites become entropically preferred, the average $n(T, x)$ increases. Recall that our parameters set modifier speciation to be 50%: $[M] / ([A] + [B]) = 0.5$. When $[B] = [A]$, both sites are entropically and enthalpically equivalent, allowing the modifiers to have the broadest probability distribution for site occupation. When $[B] \rightarrow 0$ or $[B] \rightarrow 100\%$, all modifiers are forced to associate with a specific type of network former, and hence the width of the distribution approaches a Dirac delta function.

Figure 8 shows the dependence of modifier concentration on the probability distribution. Figure 8(a) assumes $\Delta H/kT = 0$, which again represents an entropically driven system. Note that our parameters set the mole fraction of B sites to be 0.5. As $[M]$ deviates from 0.5, each modifier has fewer options for which site to occupy. As such, the standard deviation of the final system's occupied sites decreases. Figure 8(b) incorporates the enthalpic effects by modeling the modifier concentration when $\Delta H/kT = 1$. When examining Figure 8(b), refer back to the results in Figure 4, which shows that when the mole fraction of B = 0.5 and $\Delta H/kT = 1$, about 65% of the modifiers are associated with B sites. Thus, as $[M]$ increases from 0.1, more B sites have the potential to be occupied, so the number of rigid constraints in the system increases. Once $[M] > 0.5$ the modifiers run out of B sites to occupy, which causes the modifiers to occupy A sites. In the limit of $[M] = 1$, there is no competition between network formers since every site is saturated. These results demonstrate that if the enthalpic difference between site A and B approaches infinity or is inconsequential ($[M] = [A] + [B]$), the distribution collapses to a Dirac delta function [1].

## B. Temperature Dependence

Having examined the composition dependence on the number of rigid constraints per atom, we now incorporate temperature dependence. For the subsequent calculations the onset temperatures for each state are arbitrarily parameterized as follows: $T_{A,un} = 550$K, $T_{B,un} = 600$K, $T_{A,oc} = 500$K, and $T_{B,oc} = 650$K. These onset temperatures determine the localized rigidity of each constraint, the total number of rigid constraints per atom, the weighting factor (which derives from the enthalpy variance), and the resulting probability density function, as explained in Sections III and IV. Similar to our parameters from the beginning of Section V, we choose these onset temperatures so that occupying site B increases rigidity and occupying site A decreases rigidity of the localized site. The specific values of the constraint onset temperatures are chosen arbitrarily; in reality, they will vary depending on the specific glass-forming chemistry.

To incorporate the number of constraints into our investigation, we first simplify the model by assuming $w_{i,\alpha} = 1$ for all constraints $\alpha$. The distribution of the mole fraction, $N_i(x)$, was calculated previously in Figure 5. Eq. (4) and Eq. (5) from Section III, and the given onset temperatures, enable the calculation of each site's individual rigidity, $q_\alpha(T)$. Figure 9 depicts the resulting interdependence of the individual rigidities of the four states, which combine to calculate the total number of rigid constraints per atom, $n(T, x)$, of the system.

By incorporating the temperature dependence, our model no longer has a constant weighting factor or, by extension, constant $\Delta H/kT$. As $\Delta H$ increases, site B has a stronger enthalpic preference, and hence higher probability of being occupied compared to site A.

These results combine to yield the distribution of $n(T, x)$. The probability density and standard deviation for the results are presented in Figure 10. By comparing the results with the total number of constraints graphed in Figure 9, we discover that the largest distributions occur when the individual rigidities are competing for control over the macroscopic properties, with a maximum standard deviation at the total system's onset temperature. At very low or very high temperatures the standard deviation is zero, indicating that the system is in a fully rigid or fully flexible state, respectively, without any fluctuations.

Our next step in evaluating the inter-relationship of entropy and enthalpy in determining site speciation and $n(T, x)$ is to vary the quantities of each constraint, $w_\alpha$. When the quantity of each constraint increases, the system is able to have more than one rigid constraint per atom. Figure 11 maps the effect of changing the quantity of each constraint individually.

Recall from Eq. (9), if $w_{B,oc} > 1$ or $w_{A,un} > 1$, then $\Delta H$ is more negative, making site B more preferred. Conversely, if $w_{A,oc} > 1$ or $w_{B,un} > 1$, then $\Delta H$ is more positive, making site A more preferred. Our model verifies these trends through the mole fraction of sites that become occupied. Since only 50% of the sites get occupied, every modifier that occupies a B site creates a single $B, oc$ and $A, un$ site within the network. As such, $N_{B,oc} = N_{A,un}$ , $N_{A,oc} = N_{B,un}$ , and $\sum_i N_i(x) = 1$.

When the number of each type of constraint equals one, site B is preferred, yet both sites still experience modifier speciation. When $w_{B,oc}$ or $w_{A,un}$ increase, site B becomes even more preferred. When $w_{A,oc}$ or $w_{B,un}$ increase, the numbers of constraints causes site A to have more rigidity and hence become more enthalpically preferred. Therefore, the number of constraints can change the enthalpic and entropic preferences of the sites. In Figure 11, note that $n(T, x)$ for $w_{A,oc} = 2$, $w_{B,un} = 2 < w_{B,un} = 2$, $w_{B,un} = 2$. This is because when $w_{A,oc}$ or $w_{B,un}$ increase, site A more strongly dominates the total system and $q_{A,oc} + q_{B,un} < q_{B,oc} + q_{A,un}$.

Figure 11 changes the occupied and unoccupied sites individually; however, in reality, these sites occur in pairs, as explained for Eq. (9). As shown in Figure 12, when the quantities of multiple constraints change simultaneously we notice a similar effect. When $w_{A,oc} = w_{B,un} = 2$ the mole fractions $N_{A,oc} + N_{B,un} = 1$, meaning that all sites are A occupied, and equivalently B unoccupied. Since occupying site A is 100% preferred over occupying site B, increasing the number of either constraint does not change the probability of site speciation, and therefore the onset temperature remains constant. However, increasing the quantity of constraints does increase the maximum number of rigid constraints in the system.

## VI. Conclusions

Using statistical mechanics, we calculated the mole fraction distributions of site speciation which, when linked with topological constraint theory, allowed us to incorporate the contributions of localized fluctuations in the degrees of freedom in the network. When examining the composition of a desired glass network, the enthalpy differences among bonding sites, localized rigidities, concentrations of each site, and concentration of modifiers all interact to determine the probability of site speciation within the glass network. By incorporating both composition and temperature dependence into the model, we calculated the full probability density function of $n(T, x)$. We also performed several example calculations of $n(T, x)$, specifically its dependence on the quantity of each constraint, where we found that changing the quantity of each constraint can alter the resulting enthalpic and entropic preferences of the site occupation. While this paper focused on describing topological fluctuations in the liquid state, future work will investigate conditions outside of metastable equilibrium, where the structural and topological fluctuations have memory of the thermal history experienced by the system.

**Acknowledgement:** We would like to thank Sushmit Goyal of Corning Incorporated for a series of helpful discussions on this topic.

**FIGURES**

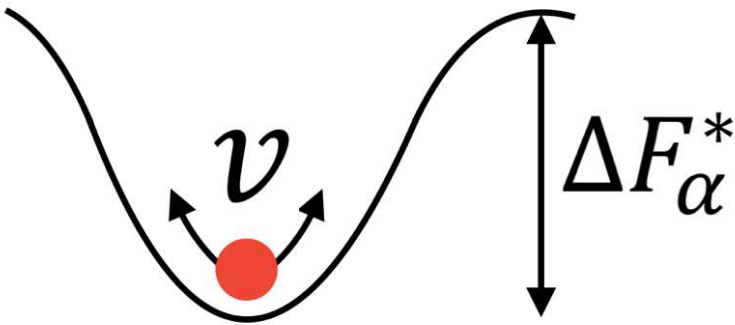

**FIG. 1.** Bond rigidity modeled as a particle in a well. Here $v$ is the vibrational attempt frequency and $\Delta F_\alpha^*$ is the activation free energy for breaking the $\alpha$ constraint. If the particle successfully escapes the well, the bond is broken and considered flexible.

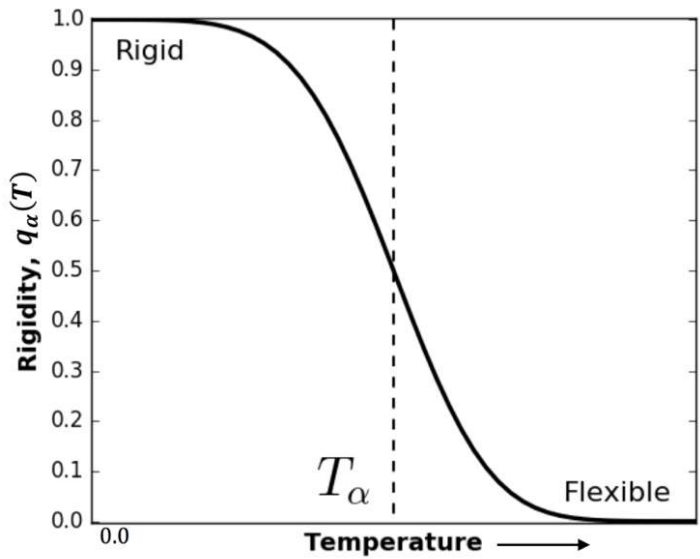

**FIG. 2.** The rigidity of constraint α as a function of temperature. The constraint onset temperature, $T_\alpha$, is the temperature at which the probability of breaking the constraint is 50% of the maximum allowed rigidity.

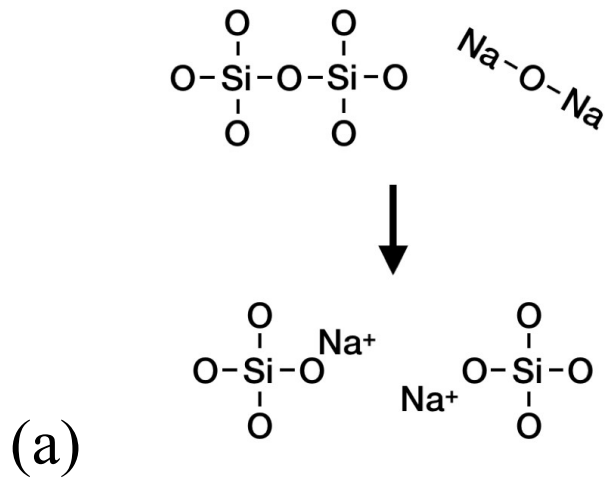

(a)

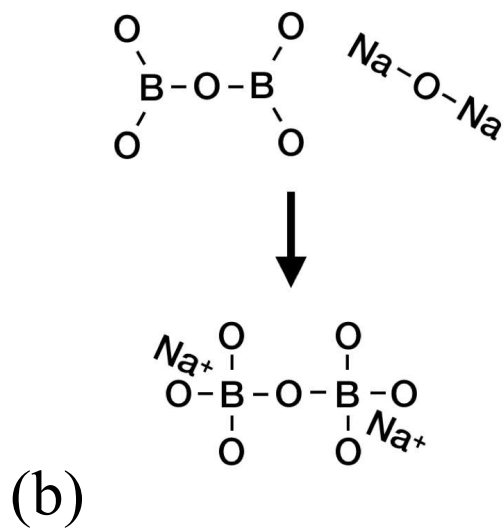

(b)

**FIG. 3.** Two ways in which a modifier can affect a network former site, e.g., in a sodium borosilicate system: (a) Forming a non-bridging oxygen, which decreases the number of rigid constraints; or (b) conversion of boron from threefold to fourfold coordination, which increases the number of rigid constraints per atom.

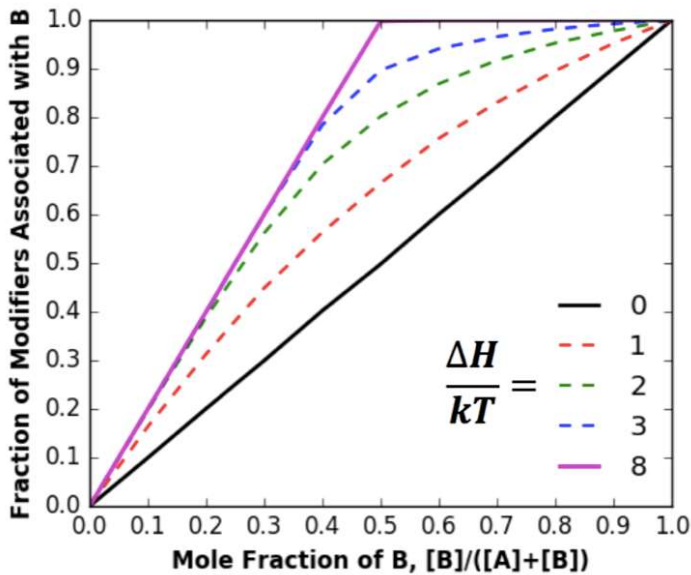

**FIG. 4.** As $\Delta H/kT \rightarrow 0$, each site is enthalpically equivalent, allowing entropy to dominate site speciation, i.e., if the mole fraction of B sites is 0.3, 30% of the modifiers associate with B. As $\Delta H/kT \rightarrow \infty$, enthalpy completely governs the system. Our model considers $H_{A,oc} > H_{B,oc}$, making site B energetically favored. Hence, when $\Delta H/kT = 8$ all of available B sites become occupied. Between the two extremes of $\Delta H/kT \rightarrow 0$ and $\Delta H/kT \rightarrow \infty$ there is a nonlinear relationship due to the competition between entropy and enthalpy (the so-called "mixed network former" effect).

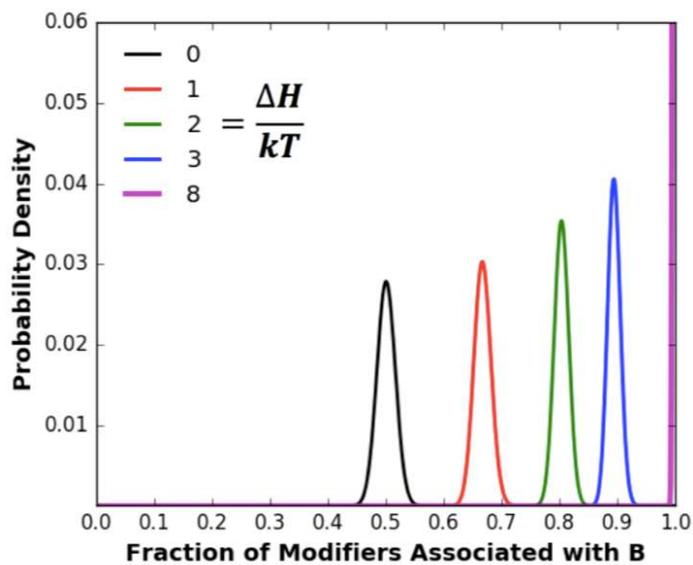

**FIG. 5.** The distribution of modifiers associated with site B, calculated for five different values of $\Delta H/kT$. As $\Delta H/kT \to 0$ the distribution follows a central hypergeometric distribution, while as $\Delta H/kT \to \infty$, it collapses to a Dirac delta function.

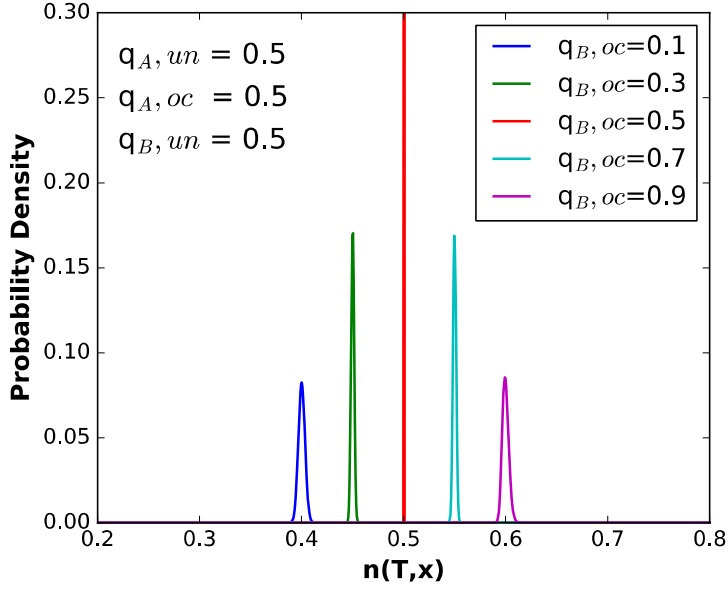

(a)

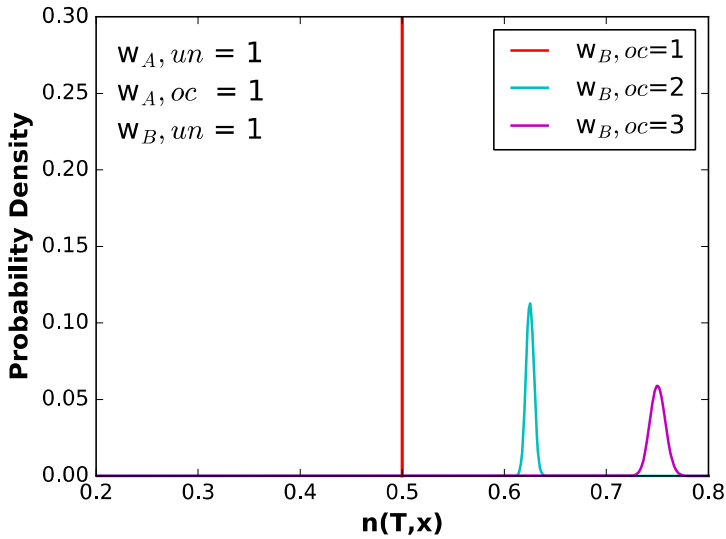

(b)

**FIG. 6.** (a) The effect of changing the magnitude of the rigidity of a localized constraint. $q_{A,un} = q_{A,oc} = q_{B,un} = 0.5$ while $q_{B,oc}$ varies over five magnitudes of rigidity. When $q_{B,oc} = 0.5$ all sites are equivalent, so each resulting network is identical, shown by the Dirac delta function. As the difference between each localized constraint increases, the breath of the distribution of possible resulting networks increases. (b) The effect of changing the quantity of each constraint. Each site has the same rigidity ($q_{A,un} = q_{A,oc} = q_{B,un} = q_{B,oc} = 0.5$) and is enthalpically and entropically identical. $w_{A,un} = w_{A,oc} = w_{B,un} = 1$ while $w_{B,oc}$ varies over three levels. The number of constraints must be a positive integer. Recall from Eq. (3) that $n(T, x)$ is the summation $\sum_\alpha w_{i,\alpha} q_\alpha(T)$. Therefore, as either $w_{i,\alpha}$ or $q_\alpha(T)$ increase, $n(T, x)$ increases.

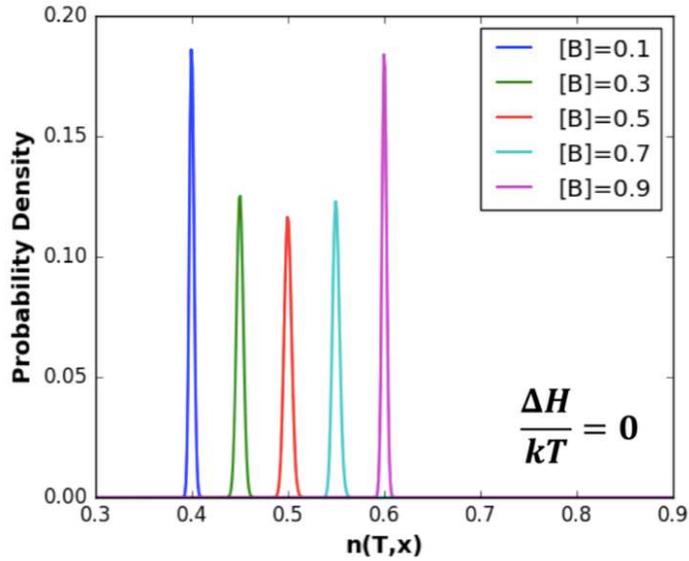

**FIG. 7.** The effect of site B concentration on $n(T, x)$. Here [B] = [B] / [A+B] and [A+B] = 1. This model assumes that $\Delta H/kT = 0$, making it a fully entropically driven system. Our parameters dictate that if a B site is occupied, it increases the number of rigid constraints per atom in the system. Hence, the graph shows that as B sites become entropically preferred, the average $n(T, x)$ increases. When $[B] \to 0$ or $[B] \to 100\%$ , the modifiers are forced to occupy either A or B sites, and hence the distribution approaches a Dirac delta function.

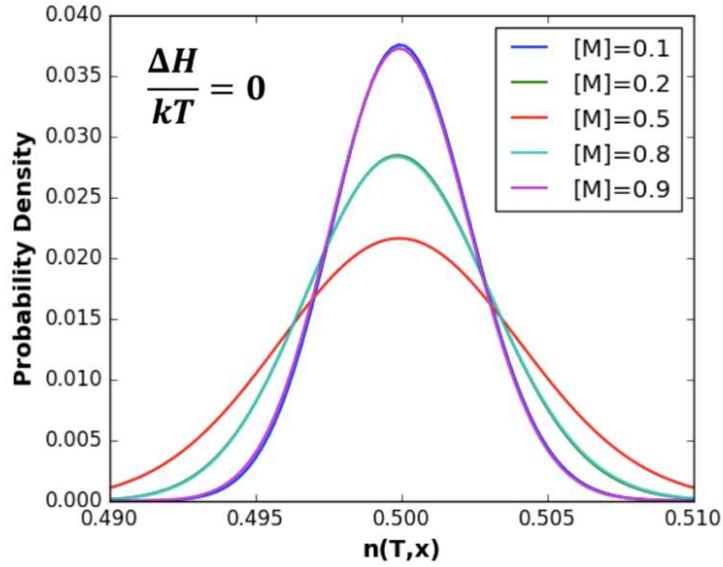

(a)

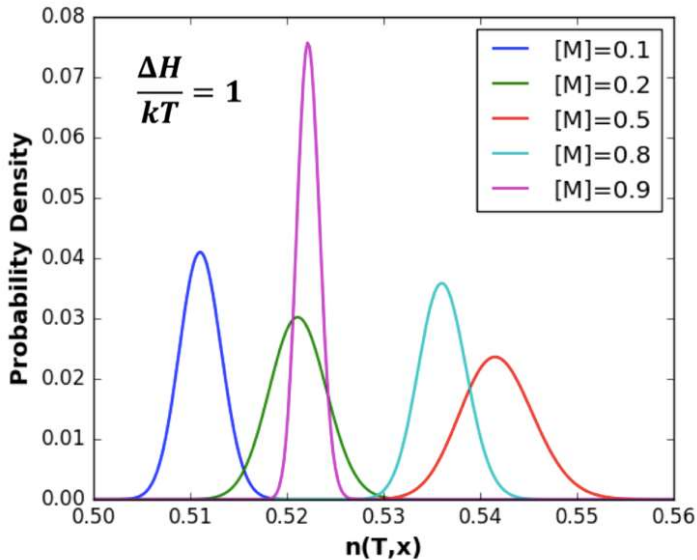

(b)

**FIG. 8.** (a) The effect of modifier concentration on $n(T, x)$. Here $[M] = [M]/([A] + [B])$ and $\Delta H/kT = 0$, meaning it is a fully entropically driven system. Since $[A] = [B] = 0.5$, when $[M] = 0.5$ the probability of occupying site A or site B is equivalent. As such, the system has the greatest standard deviation of modifier speciation. The spread of the distribution decreases as $[M] \rightarrow 0$ or $[M] \rightarrow 100\%$ since each site is occupied regardless of their relative probabilities.

Because $\Delta H/kT = 0$, the lack of site preference causes the similar results for $[M] = 0.2$ and $[M] = 0.8$, and also for $[M] = 0.1$ and $[M] = 0.9$. (b) Here we incorporate the effect of enthalpy by setting $\Delta H/kT = 1$, thereby increasing the probability of modifiers occupying site B. Starting at $[M] = 0.1$, as $[M]$ increases, more modifiers are available to occupy B sites, thereby increasing the mean and width of the distribution for $n(T, x)$. Once $[M] > 0.5$ of the total population, the modifiers run out of B sites to occupy, which forces the modifiers to occupy A sites. Paired with the fact that A occupied sites have a lower rigidity, the width and mean of the distribution decreases. As the limit of $[M] \rightarrow 1$ all sites are occupied regardless of the competition between entropic and enthalpy.

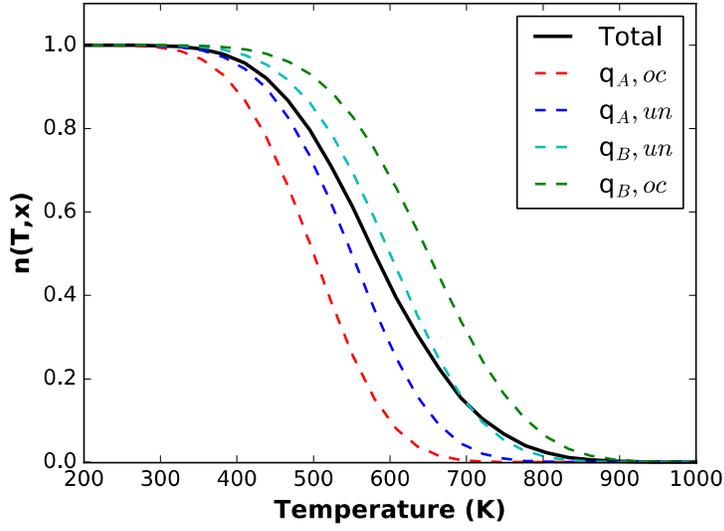

**FIG. 9.** The individual bond rigidities, $q_\alpha(T)$, and "Total" rigidity of the system (i.e., the mean number of rigid constraints per atom, $n(T, x)$) as functions of temperature. The constraint onset temperatures are: $T_{A,oc} = 500\text{K}$, $T_{A,un} = 550\text{K}$, $T_{B,un} = 600\text{K}$, and $T_{B,oc} = 650\text{K}$.

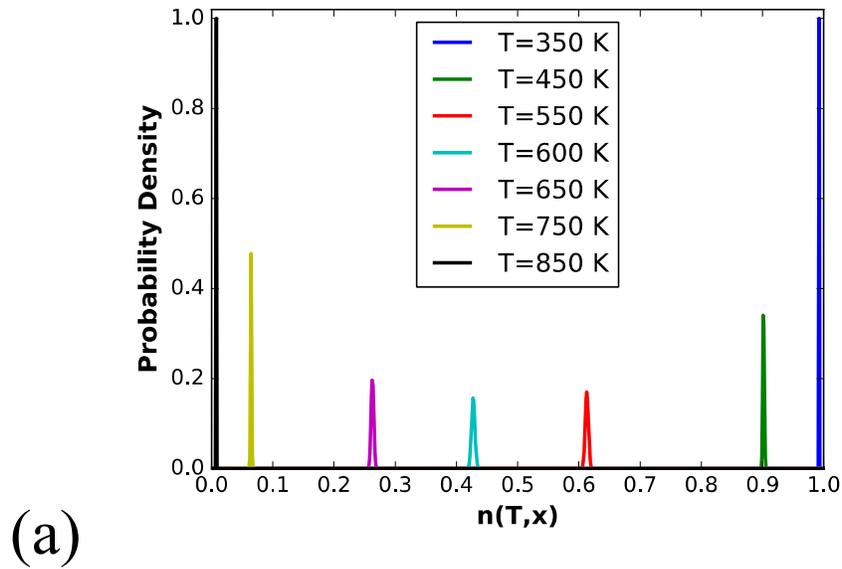

(a)

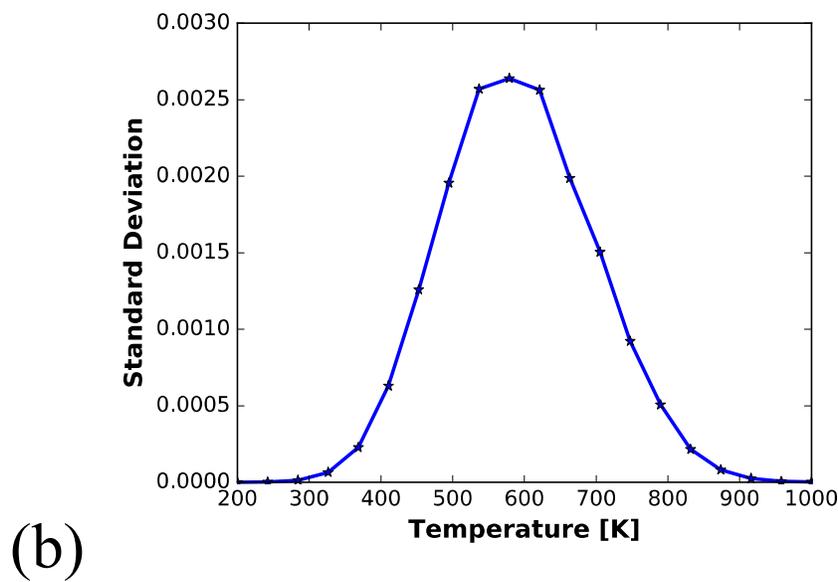

(b)

**FIG. 10.** (a) The distribution for the number of rigid constraints per atom of the total system. The width of the distribution is minimal at the extremely low or high temperatures, where the system if fully rigid or flexible. In between these extremes, the interplay of entropy and enthalpy allows the width of the distribution to increase. (b) The same results, plotting the standard deviation of the probability density function.

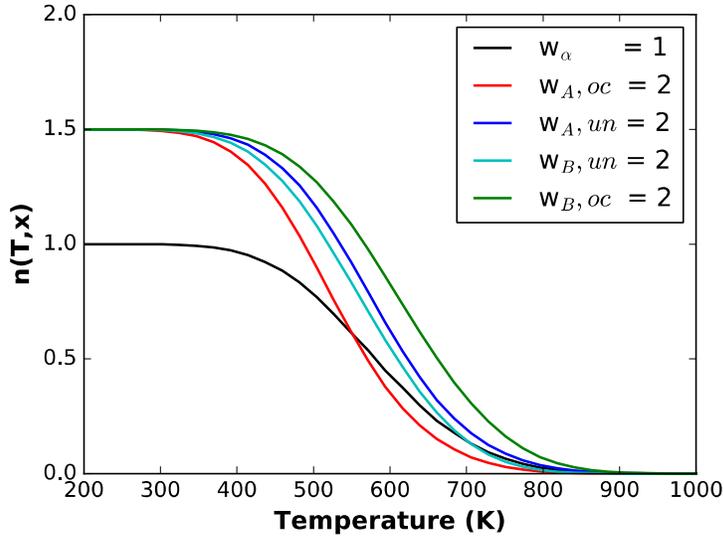

**FIG. 11.** The effect of changing the quantity of individual constraints. Here $w_\alpha = 1$ means that all constraints equal 1: $w_{A,oc} = w_{A,un} = w_{B,un} = w_{B,oc} = 1$. For each of the other cases, the specified constraint increases to 2, while the other three constraints remain equal to 1. Given Eq. (3), for each case, the maximum $n(T,x) = 0.5(2 + 1) = 1.5$, while the minimum remains 0. The onset temperature of $n(T,x)$ for increasing $w_{A,oc}$ or $w_{B,un}$ is less than when $w_{A,un}$ or $w_{B,oc}$ increases because in the first two cases site A is preferred, which has a lower comparative constraint rigidity associated with its occupation.

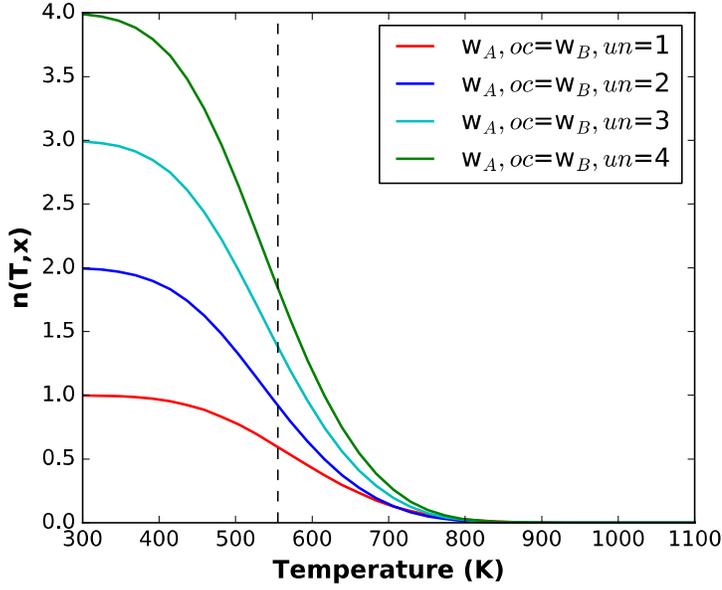

**(a)**

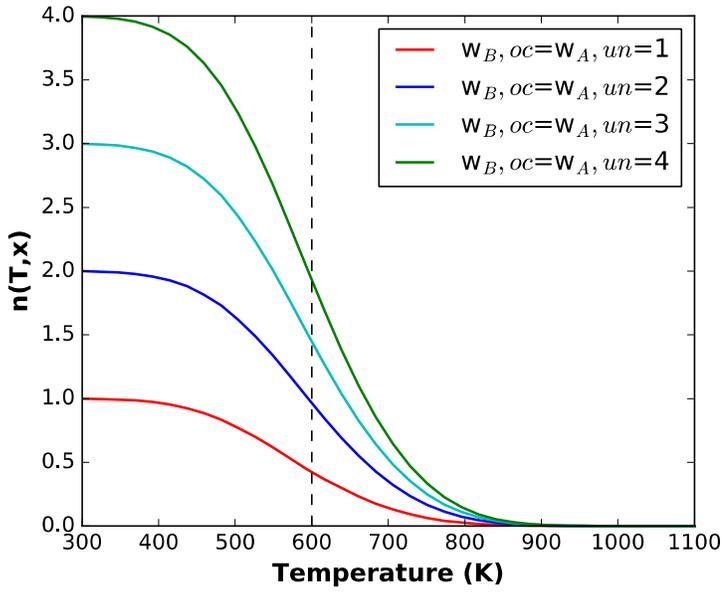

**(b)**

**FIG. 12.** (a) The effect of simultaneously changing the number of constraints, $w_{A,oc}$ and $w_{B,un}$, while $w_{B,oc} = w_{A,un} = 1$. The vertical dotted line shows the onset temperature: the temperature at which $n(T, x)$ reaches 50% of its maximum value. When the A occupied sites become more rigid than the B sites, A sites become enthalpically preferred. In fact, in our model when $w_{A,oc} = w_{B,un} = 2$ the mole fractions satisfy $N_{A,oc} + N_{B,un} = 1$, meaning that all sites are A occupied, and equivalently B unoccupied. Since site A is 100% preferred over site B, increasing the quantity of either constraint does not change the probability of site speciation; however, it does dictate the maximum $n(T, x)$. (b) Same results as (a) except shown for increasing $w_{B,oc}$ and $w_{A,un}$ simultaneously.